\documentclass[10pt,a4paper]{article}
\usepackage{graphicx}
\DeclareGraphicsExtensions{.jpg, .pdf, .png}%%% nou
\DeclareGraphicsExtensions{.jpg, .pdf, .png}%%% nou

\begin{document}

\title{AdS. Klein-Gordon equation\\
{}}
\author{Ll. Bel}

\maketitle

\abstract

I propose a generalization of the Klein-Gordon equation in the framework of AdS space-time and exhibit a four parameter family of solutions among which there is a two parameter family of time-dependent bound states.
\vspace{1cm}

{\it Introduction}
\vspace{1cm}

In 1973 E. Alvarez and I, \cite{AB}, suggested that the so-called expansion of the Universe could be due to a decreasing of the so called "speed of light constant $c$", quantified by the very simple formula:

\begin{equation}
\label{c}
\frac{\dot c}{c}=-H
\end{equation}
$H$ being the so called "Hubble constant".
This corresponds to a decreasing of $c$ by $10^{-8} m/s$ in an interval of time greater than a century, not directly observable, but it gives a meaning to start with establishing a relationship between two quantities that both depend on time, escaping thus to the apparently solid argument that only dimensionless fundamental constants could depend on time.

I have personally kept developing this point of view on several occasions \cite{Bel}, \cite{Bel2}, this paper being my last effort in this direction, while others points of view,\cite{Magueijo}, \cite{Barrow}, \cite{Moffat} have also been developed and some of them severely criticized in \cite{Ellis}.
\vspace{1cm}

{\it Space-time model}
\vspace{1cm}

Using polar coordinates, let us consider the Robertson-Walker space-time model of the Universe:
\begin{equation}
\label{ds2}
ds^2=-dt^2+\frac{1}{c^2}\left(\frac{dr^2}{1-br^2}+r^2d\Omega^2\right)
\end{equation}
where $b$ is the curvature of space and $c=c(t)$ is a time dependent function such that $c_0=c(0)$ is the speed of light at the present epoch.  Using $c(t)$ as a description of the evolution of the Universe is formally strictly equivalent to using the scale factor $a(t)=c_0/c(t)$ except that in this case it looks queer to require that a dimensionless quantity as  $a(t)$ is equal to $1$ at the present epoch, while $c(t)$ having dimensions of velocity, we can always assume that $c_0=1$ using an appropriate system of units.

%%%%%%%%%%%%%%%%%%%%%%%%%%%%%%%%%%%%%%%%%%%%%%%%%%%%%%%%%%%%%%%%%
\vspace{1cm}
{\it D'Alembertian}
\vspace{1cm}

Let us consider the D'Alembertian operator corresponding to the space-time model above acting on a function $\psi(t,r,\theta,\varphi$). A straightforward calculation yields:

\begin{eqnarray}
\label{Dalambert}
&& \hspace{-2cm}\Delta_4\Psi=-\frac{\partial^2\Psi}{\partial t^2}+3\frac{\partial\ln c}{\partial t}\frac{\partial\Psi}{\partial t} \nonumber\\[2ex]
&& \hspace{-1cm}+c^2(1-br^2)\frac{\partial^2\Psi}{\partial r^2}+\frac{2c^2}{r}\left(1-\frac{3}{2}br^2\right)\frac{\partial\Psi}{\partial r} \nonumber \\ [2ex]
&&+\frac{c^2}{r^2}\left(\frac{\partial^2\Psi}{\partial\theta^2}+\frac{1}{\sin\theta^2}\frac{\partial^2\Psi}{\partial^2\varphi}
+\frac{\cos\theta}{\sin\theta}\frac{\partial\Psi}{\partial\theta}\right)
\end{eqnarray}
\vspace{1cm}

{\it Variables separation}
\vspace{1cm}

Let us assume now that $\psi$ is the following product of three functions:

\begin{equation}
\label{Psi}
\Psi=B(t)f(r)Y(\theta,\varphi)
\end{equation}

Assuming that $Y$ is an spherical harmonic, so that:

\begin{equation}
\label{LY}
LY\equiv\frac{\partial^2Y}{\partial\theta^2}+\frac{1}{\sin\theta^2}\frac{\partial^2Y}{\partial^2\varphi}
+\frac{\cos\theta}{\sin\theta}\frac{\partial Y}{\partial\theta}=-l(l+1)Y,
\end{equation}
also that $f$ is a solution of:

\begin{equation}
\label{Lf}
Lf\equiv(1-br^2)\frac{\partial^2 f}{\partial r^2}+\frac{2}{r}\left(1-\frac{3}{2}br^2\right)\frac{\partial f}{\partial r}-\frac{l(l+1)f}{r^2}=-k_1^2f
\end{equation}
where $k_1$ is a constant. And also that $B$ is a solution of:

\begin{equation}
\label{LB}
LB\equiv-\frac{\partial^2B}{\partial t^2}+3\frac{\partial\ln c}{\partial t}\frac{\partial B}{\partial t}=k_0^2c^2B
\end{equation}
where $k_0$ is another constant, by direct substitution into (\ref{Dalambert}) we get:
\vspace{1cm}

%%%%%%%%%%%%%%%%%%%%%%%%%%%%%%%%%%%%%%%%%%%%%%%%%%%%%%%%%%%%%%%%%%%%
%{\it Result}
%\vspace{1cm}

\begin{equation}
\label{Result}
\Delta_4\Psi=(k_0^2-k_1^2)c^2\Psi
\end{equation}
\vspace{1cm}
I chose the  signs of the second members of (\ref{Lf}) and (\ref{LB}) so that:

\begin{equation}
\label{Minkowski}
\Psi=e^{i(k_0ct\pm k_1r)}Y(\theta,\varphi)
\end{equation}
when $\Lambda\rightarrow 0$, and $b\rightarrow 0$.

\vspace{1cm}
%%%%%%%%%%%%%%%%%%%%%%%%%%%%%%%%%%%%%%%%%%%%%%%%%%%%%%%%%%%
{\it Solution of the radial equation}
\vspace{1cm}

Mapple16 gives right away two independent solutions of the radial equation (\ref{Lf})

\begin{eqnarray}
\label{Legendre}
&& f_1=\frac{1}{\sqrt{r}}\hbox{LegendreP}\left(-\frac12\frac{\sqrt{b}-2\sqrt{b+k_1^2}}{\sqrt{b}},l+\frac12, \sqrt{1-br^2}\right)\\
&& f_2=\frac{1}{\sqrt{r}}\hbox{LegendreQ}\left(-\frac12\frac{\sqrt{b}-2\sqrt{b+k_1^2}}{\sqrt{b}},l+\frac12, \sqrt{1-br^2}\right)
\end{eqnarray}

\vspace{1cm}
%%%%%%%%%%%%%%%%%%%%%%%%%%%%%%%%%%%%%%%%%%%%%%%%%%%%%%%%%%%%%%
{\it Bound states, l=0 or l=-1, $b<0$}
\vspace{1cm}

Let us assume now that $b\neq 0$. In this case the two independent solutions of (\ref{Lf}) are:

\begin{equation}
\label{f}
f\pm=\frac{1}{r}\left(br+\sqrt{b(br^2-1)}\right)^\alpha, \quad \alpha=\pm\sqrt{1+\frac{k_1^2}{b}}
\end{equation}
and their behavior near the origin is:

\begin{equation}
\label{r=0}
f^\pm=e^{\alpha\ln(-b)}+O(r).
\end{equation}
For $b>0$ the solution is not regular near the origin  and therefore from now on I shall assume that $b<0$.
The behavior of the solution above when $r\rightarrow \infty$ is:

\begin{equation}
\label{r=infty}
f\pm=\left(\frac{1}{2^\alpha}\frac{1}{r}+O\left(\frac{1}{r^3}\right)\right)\frac{1}{r^\alpha},
\end{equation}
so that the space integral
\begin{equation}
\label{Int}
|f|^2=4\pi\int_0^\infty \frac{f^2r^2dr}{\sqrt{1-br^2}}
\end{equation}
is finite if $\alpha>0$, i.e., if $f=f^+$ and $k_1^2<|b|$.
Any other solution has an infinite norm.
\vspace{1cm}

%%%%%%%%%%%%%%%%%%%%%%%%%%%%%%%%%%%%%%%%%%%%%%%%%%%%%%%%%%%%%%%
{\it Time dependence}
\vspace{1cm}

To discuss the equation LB, (\ref{LB}), I shall assume that $c$ is the function of $t$ describing the Anti de Sitter model (AdS) of the Universe. It has therefore maximal space-time symmetry with negative space curvature, $b<0$, and positive cosmological constant $\Lambda>0$. In particular when $c$ is a decreasing function of time it satisfies the differential equation:

\begin{equation}
\label{Ads0}
\dot c=-c\sqrt{\lambda^2-bc^2} \ \ \ \hbox{where}\ \ \ \Lambda=3\lambda^2
\end{equation}
that integrated yields:

\begin{equation}
\label{Solution}
c=\frac{\lambda}{p}\hbox{csch}\left(\lambda t+\hbox{arccsch}\left(\frac{pc_0}{\lambda}\right)\right), \quad p=\sqrt{-b}
\end{equation}
Two other useful relations can be derived from (\ref{Ads0}), namely:

\begin{equation}
\label{AdS1}
\dot c^2=\lambda^2c^2-bc^4,
\end{equation}
and:

\begin{equation}
\label{AdS2}
\ddot c=\lambda^2c-2bc^3
\end{equation}
that follows from the preceding one after derivation and simplification.

Since $c$ is a monotonous decreasing function of $t$, it is possible to consider $B$ as a function of $c$. So that $B(t)=B(c(t))$. Using (\ref{AdS1}) and (\ref{AdS2})  leads then to the consideration of the differential equation:

\begin{equation}
\label{LBc}
LB\equiv -c^2(\lambda^2-bc)\frac{\partial^2B}{\partial c^2}+c(2\lambda^2-bc^2)\frac{\partial B}{\partial c}-k_0c^2B.
\end{equation}
$c=0$ is a regular singular value and therefore the solutions of this equation admit formal series solutions:

\begin{equation}
\label{series}
B=c^s(1+a_1c+\cdots)
\end{equation}
$s$ being a solution of the indices equation:

\begin{equation}
\label{Indicial}
-s^2+3s=0
\end{equation}
so that  $s=0$ or $s=3$.

Maple16 gives the general solution of (\ref{LBc}) as a linear combination with constant coefficients of the two particular solutions.

\begin{eqnarray}
\label{Legendre}
&& B_1=c^{3/2}\hbox{LegendreP}\left(-\frac12+\sqrt{1+\frac{k_0^2}{b}},\frac32, \sqrt{1-\frac{bc^2}{\lambda^2}}\right)\\
&& B_2=c^{3/2}\hbox{LegendreQ}\left(-\frac12+\sqrt{1+\frac{k_0^2}{b}},\frac32, \sqrt{1-\frac{bc^2}{\lambda^2}}\right)
\end{eqnarray}

But since (\ref{LBc}) is real and $B_1$ and $B_2$ are complex we have in fact four real solutions of (\ref{LBc}).
The first two terms of the power series expansions of $Re(B_1)$ and $Im(B_2)$ are:

\begin{equation}
\label{series}
Im(B_2)=\frac{\pi}{2}Re(B_1)=-\frac{\sqrt{\pi}}{8}\frac{2^{3/4}}{(-\frac32\frac{b}{\Lambda})^{3/4}\Lambda\sqrt{\pi}}(2\Lambda+3k_0^2 c^2)
\end{equation}
This proves that they belong to the index $s=0$ and that they are proportional with a factor $(1/2)\pi$. Extending the series a few more terms it is easy to prove that $Im(B_1)=0$ and that  $Re(B_2)$ belongs to the index $s=3$. This distinguishes this latter function as the only one that goes to zero when $c$ goes to zero.

The function $B_2$ and its complex conjugate $\bar B_2$ can therefore be considered as the fundamental complex solution of (\ref{LBc}).

I have thus proved that there exists a system of modes:

\begin{equation}
\label{Modes}
\psi=B_2(t,k_0)f^+(r,k_1)Y_l^m(\theta,\varphi)
\end{equation}
depending on four parameters $(k_0,k_1,l,m )$ that are solutions of a generalized Klein-Gordon:

\begin{equation}
\label{KG}
\Delta_4\psi=(k_0^2-k_1^2)c^2\psi
\end{equation}
Noteworthy is the fact that with $l=0$ or $l=-1$ and $k_1^2<|b|$ the corresponding $f^+$ time-independent factor norm is finite and therefore $\psi$ in this case describes a time-dependent bound state.

A concomitant consequence to assuming that $c$ is a function of time  is that it might be necessary or plausible to consider also the time dependence of some of the other so called "fundamental constants", \cite{Barrow}, \cite{Bel}. In this latter arXiv paper I found plausible to accept that Newton´s gravitational constant $G$ and the fine structure constant $\alpha$ should be kept constants. And that on the contrary the elementary charge $e$, the electric permittivity $\epsilon$, the magnetic permeability $\mu$, the mass of the elementary particles $m$ and the Planck's constant $h$ should vary as follows:

\begin{equation}
\label{Constants}
\epsilon=\epsilon_0\frac{c_0}{c}, \ \ \mu=\mu_0\frac{c_0}{c},\ \ e=e_0\frac{c}{c_0},\ h=h_0\frac{c^2}{c_0^2}, \ e=e_0\frac{c}{c_0},
\ m=m_0\frac{c}{c_0},
\end{equation}
If this is the case then we have that:

\begin{equation}
\label{mbar}
\frac{m^2c^2}{\hbar^2}=\frac{m_0^2c_0^2}{\hbar_0^2}
\end{equation}
and (\ref{KG}) can equivalently be written:

\begin{equation}
\label{KGX}
\Delta_4\psi=\frac{m^2c^4}{\hbar^2}\psi,\ \ \ \hbox{with} \ \ \ \frac{m^2c^2}{\hbar^2}=k_0^2-k_1^2.
\end{equation}

Figure 1 is the graph of $c$ corresponding to $b=-0.45$, and $\Lambda=1.65$ (units as in \cite{Bel2}).

Figures 2 are the graphs of the real and imaginary parts of $B_2(c(t))$ with $k_0=6$.

\begin{figure}[htbp]
\begin{center}
\begin{minipage}[b]{25cm}
\includegraphics[width=8cm]{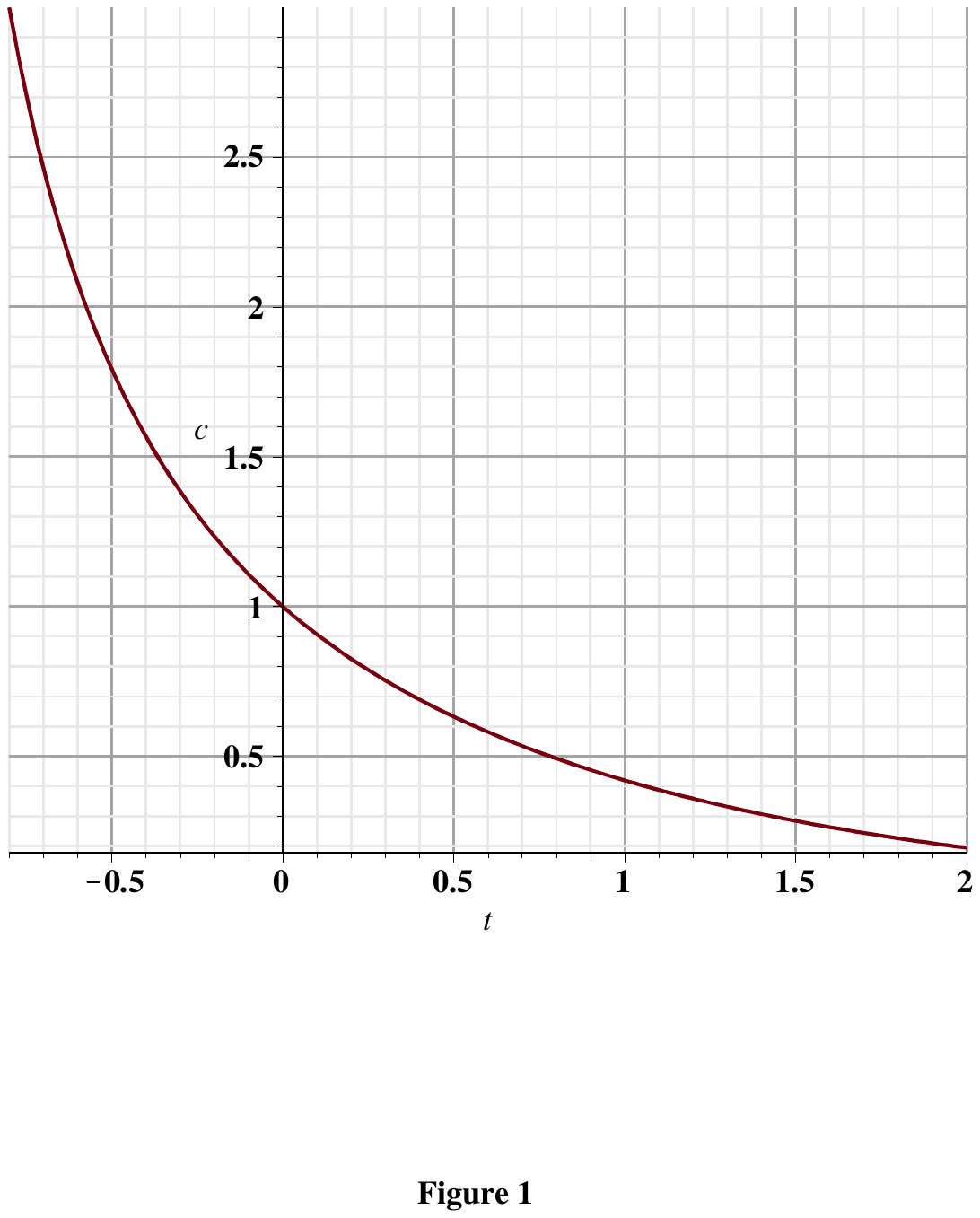}
\includegraphics[width=8cm]{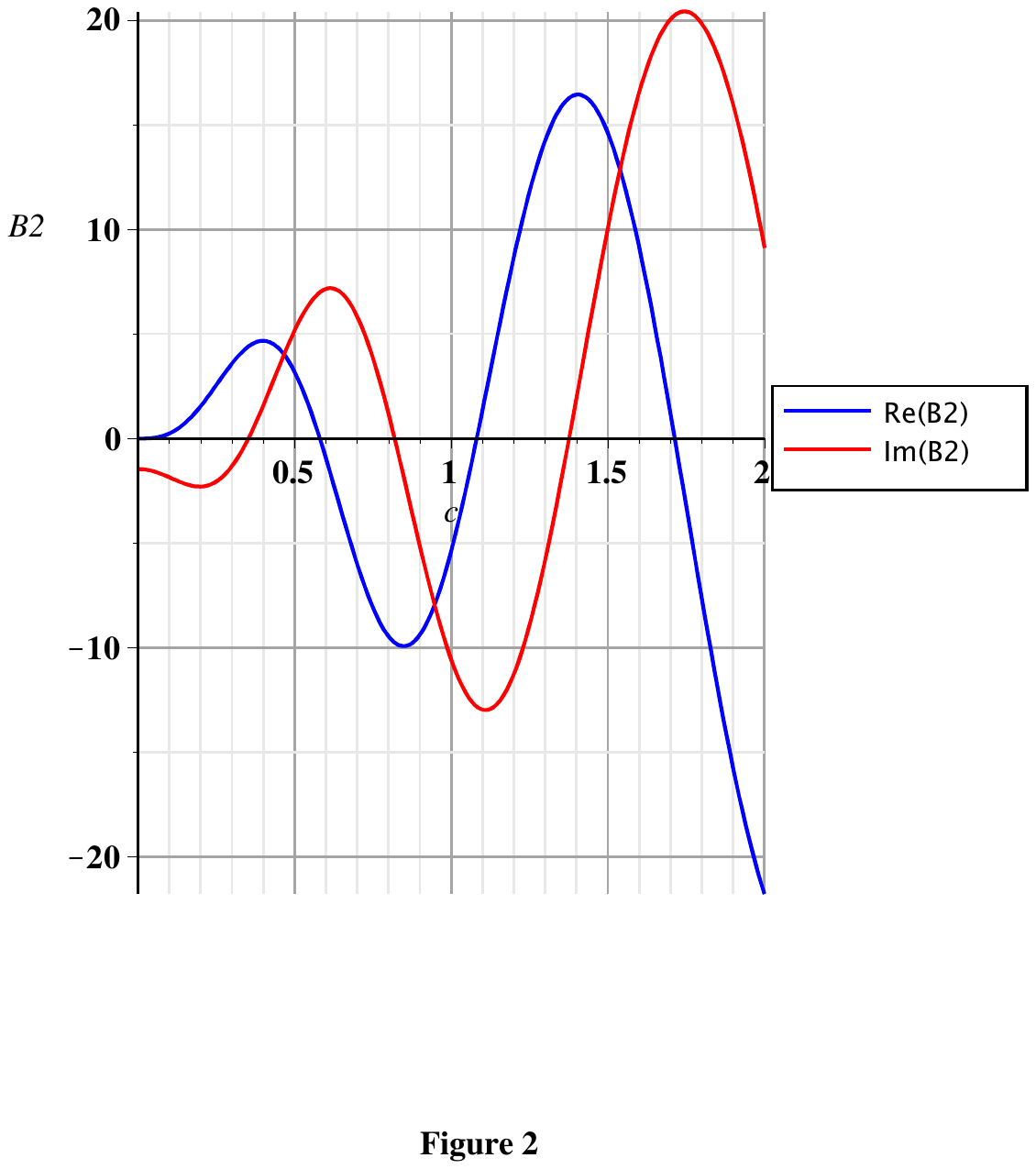}
\end{minipage}
\end{center}
\end{figure}

\end{document}